\documentclass[aps,twocolumn,floatfix,showpacs]{revtex4}
\usepackage{graphicx}
\usepackage{amsmath}
\usepackage{amssymb}
\usepackage{bm}

\begin{document}

\title{Quantized conductance at the Majorana phase transition in a disordered superconducting wire}
\author{A. R. Akhmerov}
\affiliation{Instituut-Lorentz, Universiteit Leiden, P.O. Box 9506, 2300 RA Leiden, The Netherlands}
\author{J. P. Dahlhaus}
\affiliation{Instituut-Lorentz, Universiteit Leiden, P.O. Box 9506, 2300 RA Leiden, The Netherlands}
\author{F. Hassler}
\affiliation{Instituut-Lorentz, Universiteit Leiden, P.O. Box 9506, 2300 RA Leiden, The Netherlands}
\author{M. Wimmer}
\affiliation{Instituut-Lorentz, Universiteit Leiden, P.O. Box 9506, 2300 RA Leiden, The Netherlands}
\author{C. W. J. Beenakker}
\affiliation{Instituut-Lorentz, Universiteit Leiden, P.O. Box 9506, 2300 RA Leiden, The Netherlands}

\date{September, 2010}

\begin{abstract}
Superconducting wires without time-reversal and spin-rotation symmetries can be driven into a topological phase that supports Majorana bound states. Direct detection of these zero-energy states is complicated by the proliferation of low-lying excitations in a disordered multi-mode wire. We show that the phase transition itself is signaled by a quantized thermal conductance and electrical shot noise power, irrespective of the degree of disorder. In a ring geometry, the phase transition is signaled by a period doubling of the magnetoconductance oscillations. These signatures directly follow from the identification of the sign of the determinant of the reflection matrix as a topological quantum number.
\end{abstract}

\maketitle

It has been predicted theoretically \cite{Lut10} that the \textit{s}-wave proximity effect of a superconducting substrate can drive a spin-polarized and spin-orbit coupled semiconductor nanowire into a topological phase \cite{Vol97,Rea00,Kit01}, with a Majorana fermion trapped at each end of the wire. There exists now a variety of proposals \cite{Has10,Ali10,Sau10} for topological quantum computing in nanowires that hope to benefit from the long coherence time expected for Majorana fermions. A superconducting proximity effect in InAs wires (which have the required strong spin-orbit coupling) has already been demonstrated in zero magnetic field \cite{Dam06}, and now the experimental challenge is to drive the system through the Majorana phase transition in a parallel field. 

Proposals to detect the topological phase have focused on the detection of the Majorana bound states at the end points of the wire, through their effect on the current-voltage characteristic \cite{Law09,Lin10} or the AC Josephson effect \cite{Kwo04,Fu08}. These signatures of the topological phase would stand out in a clean single-mode wire, but the multiple modes and potential fluctuations in a realistic system are expected to produce a chain of coupled Majorana's \cite{Shi10,Neu10}, which would form a band of low-lying excitations that would be difficult to distinguish from ordinary fermionic bound states \cite{Fle10}.

Here we propose an altogether different detection strategy: Rather than trying to detect the Majorana bound states inside the topological phase, we propose to detect the phase transition itself. A topological phase transition is characterized by a change in the topological quantum number $Q$. The value of $Q=(-1)^{m}$ is determined by the parity of the number $m$ of Majorana bound states at each end of the wire, with $Q=-1$ in the topological phase \cite{Hasan10}.

In accord with earlier work \cite{Mer02}, we relate the topological quantum number to  the determinant of the matrix $r$ of quasiparticle reflection amplitudes, which crosses zero at the phase transition. This immediately implies a unit transmission eigenvalue at the transition. Disorder may shift the position of the transition but it cannot affect the unit height of the transmission peak. We propose experiments to measure the transmission peak in both thermal and electrical transport properties, and support our analytical predictions by computer simulations.

We consider a two-terminal transport geometry, consisting of a disordered superconducting wire of length $L$, connected by clean normal-metal leads to reservoirs in thermal equilibrium (temperature $\tau_{0}$). The leads support $2N$ right-moving modes and $2N$ left-moving modes at the Fermi level, with mode amplitudes $\psi_{+}$ and $\psi_{-}$, respectively. The spin degree of freedom is included in the number $N$, while the factor of two counts the electron and hole degree of freedom.

The $4N\times 4N$ unitary scattering matrix ${\cal S}$ relates incoming and outgoing mode amplitudes,
\begin{equation}
\begin{pmatrix}
\psi_{-,{\rm L}}\\
\psi_{+,{\rm R}}
\end{pmatrix}
={\cal S}
\begin{pmatrix}
\psi_{+,{\rm L}}\\
\psi_{-,{\rm R}}
\end{pmatrix},\;\;
{\cal S}=\begin{pmatrix}
r&t'\\
t&r'
\end{pmatrix},\label{Sdef}
\end{equation}
where the labels L and R distinguish modes in the left and right lead. The four blocks of ${\cal S}$ define the $2N\times 2N$ reflection matrices $r,r'$ and transmission matrices $t,t'$.

Time-reversal symmetry and spin-rotation symmetry are broken in the superconductor, but electron-hole symmetry remains. At the Fermi energy electron-hole symmetry implies that if $(u,v)$ is an electron-hole eigenstate, then also $(v^{\ast},u^{\ast})$. Using this symmetry we can choose a basis such that all modes have purely real amplitudes. In this socalled Majorana basis ${\cal S}$ is a real orthogonal matrix, ${\cal S}^{\rm t}={\cal S}^{\dagger}={\cal S}^{-1}$. (The superscript $t$ indicates the transpose of a matrix.) More specifically, since ${\rm det}\,{\cal S}=1$ the scattering matrix is an element of the special orthogonal group ${\rm SO}(4N)$. This is symmetry class D \cite{Boc00,Bro00,Mot01,Gru05,Eve08,note4}.

The scattering matrix in class D has the polar decomposition
\begin{equation}
{\cal S}=\begin{pmatrix}
O_{1}&0\\
0&O_{2}
\end{pmatrix}\begin{pmatrix}
\tanh\Lambda&(\cosh\Lambda)^{-1}\\
(\cosh\Lambda)^{-1}&-\tanh\Lambda
\end{pmatrix}\begin{pmatrix}
O_{3}&0\\
0&O_{4}
\end{pmatrix},\label{SOLambda}
\end{equation}
in terms of four orthogonal matrices $O_{p}\in{\rm SO}(2N)$ and a diagonal real matrix $\Lambda$ with diagonal elements $\lambda_{n}\in(-\infty,\infty)$. The absolute value $|\lambda_{n}|$ is called a Lyapunov exponent, related to the transmission eigenvalue $T_{n}\in[0,1]$ by $T_{n}=1/\cosh^{2}\lambda_{n}$. We identify
\begin{equation}
Q={\rm sign}\,{\cal Q},\;\;{\cal Q}={\rm Det}\,r={\rm Det}\,r'={\prod_{n=1}^{2N}}\tanh\lambda_{n}.\label{Qdef}
\end{equation}
This relation expresses the fact that reflection from a Majorana bound state contributes a scattering phase shift of $\pi$, so a phase factor of $-1$. The sign of $\prod_{n}\tanh\lambda_{n}$ thus equals the parity of the number $m$ of Majorana bound states at one end of the wire \cite{note1}. (It makes no difference which end, and indeed $r$ and $r'$ give the same $Q$.) 

To put this expression for $Q$ into context, we first note that it may be written equivalently as $Q={\rm Det}\,O_{1}O_{3}$ if we restrict the $\lambda_{n}$'s to non-negative values and allow ${\rm Det}\,O_{p}$ to equal either $+1$ or $-1$. The sign of $Q$ then corresponds to the topological classification of a class-D network model derived by Merz and Chalker \cite{Mer02}. We also note that $Q$ can be written equivalently in terms of the Pfaffian of $\ln{\cal M}{\cal M}^{\dagger}$ (with ${\cal M}$ the transfer matrix in a suitable basis) \cite{note1}. A Pfaffian relation for the topological quantum number $Q_{\rm clean}$ in class D has been derived by Kitaev \cite{Kit01} for a clean, translationally invariant system. We will verify later on that $Q$ and $Q_{\rm clean}$ agree for a clean system.

An immediate consequence of Eq.\ \eqref{Qdef} is that at the topological phase transition one of the $\lambda_{n}$'s vanishes \cite{Mer02,Mot01,Gru05}, so the corresponding transmission eigenvalue $T_{n}=1$ at the transition point. The sign change of $Q$ ensures that $T_{n}$ fully reaches its maximal value of unity, it cannot stop short of it without introducing a discontinuity in ${\cal Q}$. Generically there will be only a single unit transmission eigenvalue at the transition, the others being exponentially suppressed by the superconducting gap. The thermal conductance $G_{\rm th}=G_{0}\sum_{n}T_{n}$ of the wire will then show a peak of quantized height $G_{0}=\pi^{2}k_{B}^{2}\tau_{0}/6h$ at the transition.

Our claim of a quantized conductance at the transition point is consistent with earlier work \cite{Bro00,Mot01,Gru05,Eve08} on class D ensembles. There a broad distribution of the conductance was found in the large-$L$ limit, but the key difference is that we are considering a single disordered sample of finite length, and the value of the control parameter at which the conductance is quantized is sample specific. We will now demonstrate how the peak of quantized conductance arises, first for a simple analytically solvable model, then for a more complete microscopic Hamiltonian that we solve numerically.

The analytically solvable model is the effective low-energy Hamiltonian of a class-D superconductor with a random gap, which for a single mode in the Majorana basis has the form
\begin{equation}
H=-i\hbar v_{F}\sigma_{z}\partial/\partial x+\Delta(x)\sigma_{y}.\label{Hdef}
\end{equation}
We have assumed, for simplicity, that right-movers and left-movers have the same velocity $v_{F}$, but otherwise this is the generic form to linear order in momentum, constrained by the electron-hole symmetry requirement $H=-H^{\ast}$. An eigenstate $\Psi$ of $H$ at energy zero satisfies
\begin{equation}
\Psi(x)=\exp\left(-\frac{1}{\hbar v_{F}}\sigma_{x}\int_{0}^{x}\,\Delta(x')dx'\right)\Psi(0).\label{Psix}
\end{equation}
By substituting $\Psi(0)=(1,r)$, $\Psi(L)=(t,0)$ we obtain the reflection amplitude
\begin{equation}
r=\tanh(L\bar{\Delta}/\hbar v_{F}),\;\;\bar{\Delta}=L^{-1}\int_{0}^{L}\,\Delta(x)dx.\label{rbarDelta}
\end{equation}
In this simple model, a change of sign of the spatially averaged gap $\bar{\Delta}$ is the signature of a topological phase transition \cite{note2}.

If $\bar{\Delta}$ is varied by some external control parameter, the thermal conductance $G_{\rm th}=G_{0}\cosh^{-2}(L\bar{\Delta}/\hbar v_{F})$ has a peak at the transition point $\bar{\Delta}=0$, of height $G_{0}$ and width $\hbar v_{F}/L$ (Thouless energy). The $1/\cosh^{2}$ line shape is the same as for a thermally broadened tunneling resonance, but the quantized peak height (irrespective of any asymmetry in the coupling to the left and right lead) is highly distinctive.

\begin{figure}[tb]
\centerline{\includegraphics[width=0.8\linewidth]{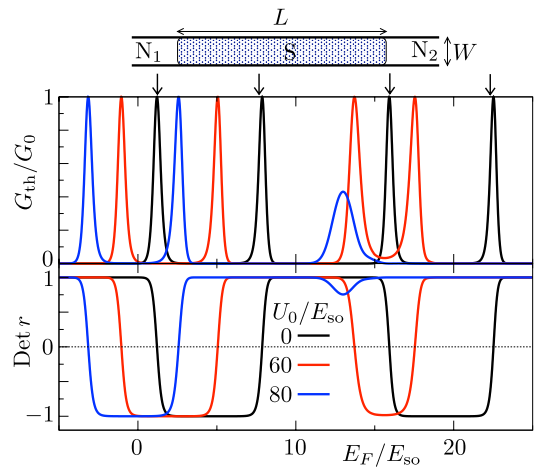}}
\caption{\label{fig_Gthermal}
Thermal conductance and determinant of reflection matrix of a disordered multimode superconducting wire as a function of Fermi energy. The curves are calculated numerically from the Hamiltonian \eqref{HBdG}--\eqref{HRashba} on a square lattice (lattice constant $a=l_{\rm so}/20$), for parameter values $W=l_{\rm so}$, $L=10\,l_{\rm so}$, $\Delta=10\,E_{\rm so}$, $g_{\rm eff}\mu_{B}B=21\,E_{\rm so}$, and three different disorder strengths $U_{0}$. The arrows indicate the expected position of the topological phase transition in an infinite clean wire ($U_{0}=0$, $L\rightarrow\infty$), calculated from Eq.\ \eqref{Qclean}. Disorder reduces the topologically nontrivial interval (where ${\rm Det}\,r<0$), and may even remove it completely, but the conductance quantization remains unaffected as long as the phase transition persists.
}
\end{figure}

For a more realistic microscopic description of the quantized conductance peak, we have performed a numerical simulation of the model \cite{Lut10} of a semiconductor nanowire on a superconducting substrate. The Bogoliubov-De Gennes Hamiltonian
\begin{equation}
{\cal H}=\begin{pmatrix}
H_{\rm R}-E_{F}&\Delta\\
\Delta^{\ast}&E_{F}-\sigma_{y}H_{\rm R}^{\ast}\sigma_{y}
\end{pmatrix}\label{HBdG}
\end{equation}
couples electron and hole excitations near the Fermi energy $E_{F}$ through an \textit{s}-wave superconducting order parameter $\Delta$. Electron-hole symmetry is expressed by
\begin{equation}
\sigma_{y}\tau_{y}{\cal H}^{\ast}\sigma_{y}\tau_{y}=-{\cal H},\label{ehsymmetry}
\end{equation}
where the Pauli matrices $\sigma_{y}$ and $\tau_{y}$ act, respectively, on the spin and the electron-hole degree of freedom. The excitations are confined to a wire of width $W$ and length $L$ in the $x-y$ plane of the semiconductor surface inversion layer, where their dynamics is governed by the Rashba Hamiltonian
\begin{equation}
H_{\rm R}=\frac{\bm{p}^{2}}{2m_{\rm eff}}+U(\bm{r})+\frac{\alpha_{\rm so}}{\hbar}(\sigma_{x}p_{y}-\sigma_{y}p_{x})+\tfrac{1}{2}g_{\rm eff}\mu_{B}B\sigma_{x}.\label{HRashba}
\end{equation}
The spin is coupled to the momentum $\bm{p}=-i\hbar\partial/\partial{\bm r}$ by the Rashba effect, and polarized through the Zeeman effect by a magnetic field $B$ parallel to the wire (in the $x$-direction). Characteristic length and energy scales are $l_{\rm so}=\hbar^{2}/m_{\rm eff}\alpha_{\rm so}$ and $E_{\rm so}=m_{\rm eff}\alpha_{\rm so}^{2}/\hbar^{2}$. Typical values in InAs are $l_{\rm so}=100\,{\rm nm}$, $E_{\rm so}=0.1\,{\rm meV}$, $g_{\rm eff}\mu_{B}=2\,{\rm meV}/{\rm T}$.

We have solved the scattering problem numerically \cite{Wim09} by discretizing the Hamiltonian \eqref{HBdG} on a square lattice (lattice constant $a$), with a short-range electrostatic disorder potential $U(x,y)$ that varies randomly from site to site, distributed uniformly in the interval $(-U_{0},U_{0})$. (Equivalent results are obtained for long-range disorder \cite{note1}.) The disordered superconducting wire (S) is connected at the two ends to clean metal leads (${\rm N}_{1},{\rm N}_{2}$), obtained by setting $U\equiv 0$, $\Delta\equiv 0$ for $x<0$, $x>L$. Results for the thermal conductance and topological quantum number are shown in Fig.\ \ref{fig_Gthermal}, as a function of the Fermi energy (corresponding to a variation in gate voltage). For the parameters listed in the caption the number $N$ of modes in the normal leads increases from 1 to 2 at $E_{F}/E_{\rm so}\approx 10$ and from 2 to 3 at $E_{F}/E_{\rm so}\approx 15$. We emphasize that Fig.\ \ref{fig_Gthermal} shows raw data, without any averaging over disorder.

For a clean system ($U_{0}=0$, black curves) the results are entirely as expected: A topologically nontrivial phase (with ${\rm Det}\,r<0$) may appear for odd $N$ while there is no topological phase for $N$ even \cite{Wim10,Lut10b,Pot10}. The topological quantum number of an infinitely long clean wire (when the component $p_{x}$ of momentum along the wire is a good quantum number) can be calculated from the Hamiltonian ${\cal H}(p_{x})$ using Kitaev's Pfaffian formula \cite{Kit01,Lut10b},
\begin{equation}
Q_{\rm clean}={\rm sign}\,\bigl({\rm Pf}\,[\sigma_{y}\tau_{y}H(0)]{\rm Pf}\,[\sigma_{y}\tau_{y}H(\pi/a)]\bigr).\label{Qclean}
\end{equation}
(The multiplication by $\sigma_{y}\tau_{y}$ ensures that the Pfaffian is calculated of an antisymmetric matrix.) The arrows in Fig.\ \ref{fig_Gthermal} indicate where $Q_{\rm clean}$ changes sign, in good agreement with the sign change of $Q$ calculated from Eq.\ \eqref{Qdef}. (The agreement is not exact because $L$ is finite.)  

Upon adding disorder $Q_{\rm clean}$ can no longer be used (because $p_{x}$ is no longer conserved), and we rely on a sign change of $Q$ to locate the topological phase transition. Fig.\ \ref{fig_Gthermal} shows that disorder moves the peaks closer together, until they merge and the topological phase disappears for sufficiently strong disorder. We have also observed the inverse process, a disorder-induced splitting of a peak and the appearance of a topological phase, in a different parameter regime than shown in Fig.\ \ref{fig_Gthermal}. Our key point is that, as long as the phase transition persists, disorder has no effect on the height of the conductance peak, which remains precisely quantized --- without any finite-size effects.

Since electrical conduction is somewhat easier to measure than thermal conduction, we now discuss two alternative signatures of the topological phase transition which are purely electrical. An electrical current $I_{1}$ is injected into the superconducting wire from the normal metal contact ${\rm N}_{1}$, which is at a voltage $V_{1}$ relative to the grounded superconductor. An electrical current $I_{2}$ is transmitted as quasiparticles into the grounded contact ${\rm N}_{2}$, the difference $I_{1}-I_{2}$ being drained to ground as Cooper pairs via the superconductor. The nonlocal conductance $G=\bar{I}_{2}/V_{1}$ is determined by the time averaged current $\bar{I}_{2}$, while the correlator of the time dependent fluctuations $\delta I_{2}$ determines the shot noise power $P=\int_{-\infty}^{\infty}dt\,\langle\delta I_{2}(0)\delta I_{2}(t)\rangle$ (in the regime $k_{B}\tau_{0}\ll eV_{1}$ where thermal noise can be neglected). 

\begin{figure}[tb]
\centerline{\includegraphics[width=0.8\linewidth]{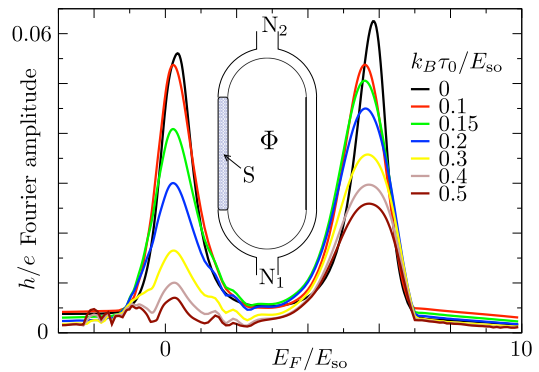}}
\caption{\label{fig_GAB}
Fourier amplitude with flux periodicity $h/e$ of the magnetoconductance oscillations, calculated numerically from the Hamiltonian \eqref{HBdG}--\eqref{HRashba} for a single disorder strength $U_{0}=50\,E_{\rm so}$ and seven different temperatures $\tau_{0}$. The inset shows the Aharonov-Bohm ring geometry. The parameters of the superconducting segment of the ring (S) are the same as in Fig.\ \ref{fig_Gthermal}, with $N=1$ in this range of Fermi energies. The normal part of the ring has $N=8$ propagating modes to avoid localization by the disorder (which has the same strength throughout the ring).
}
\end{figure}

These two electrical transport properties are given in terms of the $N\times N$ transmission matrices $t_{ee}$ and $t_{he}$ (from electron to electron and from electron to hole) by the expressions \cite{Ana96}
\begin{align}
&G=(e^{2}/h)\,{\rm Tr}\,{\cal T}_{-},\;\;P=(e^{3}V_{1}/h)\,{\rm Tr}\,\bigl({\cal T}_{+}-{\cal T}_{-}^{2}\bigr),\label{GPdef}\\
&{\cal T}_\pm=t_{ee}^{\dagger}t_{ee}^{\vphantom{\dagger}}\pm t_{he}^{\dagger}t_{he}^{\vphantom{\dagger}}.\label{Tpmdef}
\end{align}
Electron-hole symmetry relates $t_{ee}=t_{hh}^{\ast}$ and $t_{he}=t_{eh}^{\ast}$. This directly implies that ${\rm Tr}\,{\cal T}_{+}=\frac{1}{2}{\rm Tr}\,tt^{\dagger}=\frac{1}{2}\sum_{n}T_{n}$. If in addition we assume that at most one of the $T_{n}$'s is nonzero we find that ${\cal T}_{-}$ vanishes \cite{note1}. We conclude that $G$ remains zero across the topological phase transition, while $P/V_{1}$ peaks at the quantized value $e^{3}/2h$. This is the second signature of the phase transition \cite{note3}.

The third signature is in the electrical conductance. Since $G=0$ for a single open transmission channel, we add (topologically trivial) open channels by means of a parallel normal metal conductor in a ring geometry. A magnetic flux $\Phi$ through the ring produces Aharonov-Bohm oscillations with a periodicity $\Delta\Phi=h/e^{\ast}$. The effective charge $e^{\ast}=e$ if electrons or holes can be transmitted individually through the superconducting arm of the ring, while $e^{\ast}=2e$ if only Cooper pairs can be transmitted \cite{But86,Ben10}. We thus expect a period doubling from $h/2e$ to $h/e$ of the magnetoconductance oscillations at the phase transition, which is indeed observed in the computer simulations (Fig.\ \ref{fig_GAB}). To show the relative robustness of the effect to thermal averaging, we repeated the calculation at several different temperatures $\tau_{0}$. For $E_{\rm so}\simeq 0.1\,{\rm meV}$ the characteristic peak at the phase transition remains visible for temperatures in the readily accessible range of 100--500~mK.

In conclusion, our analytical considerations and numerical simulations of a model Hamiltonian \cite{Lut10} of a disordered InAs wire on a superconducting substrate show three signatures of the transition into the topological phase (Figs.\ \ref{fig_Gthermal} and \ref{fig_GAB}): A quantized thermal conductance and electrical shot noise \cite{note3}, and a period doubling of the magnetoconductance oscillations. These unique signatures of the Majorana phase transition provide alternatives to the detection of Majorana bound states \cite{Law09,Lin10,Kwo04,Fu08,Shi10,Fle10}, which are fundamentally insensitive to the obscuring effects of disorder in a multimode wire.

We thank N. Read for alerting us to relevant literature. This research was supported by the Dutch Science Foundation NWO/FOM, by the Deutscher Akademischer Austausch Dienst DAAD, and by an ERC Advanced Investigator Grant.

\appendix
\section{Derivation of the scattering formula for the topological quantum number}
\label{AppQ}

\subsection{Pfaffian form of the topological quantum number}
\label{DetPf}

The topological quantum number $Q$ of a disordered wire is given in Eq.\ \eqref{Qdef} as the sign of the determinant of the reflection matrix. That is the form which is most convenient for computations. In order to derive this relationship and also to compare it with results in the literature for translationally invariant systems \cite{Kit01}, it is convenient to rewrite it in terms of the transfer matrix ${\cal M}$. It then takes the form of a Pfaffian, rather than a determinantal, relation.

The $4N\times 4N$ transfer matrix ${\cal M}$ relates the mode amplitudes to the right (R) and to the left (L) of the disordered wire,
\begin{equation}
\begin{pmatrix}
\psi_{+,{\rm R}}\\
\psi_{-,{\rm R}}
\end{pmatrix}={\cal M}
\begin{pmatrix}
\psi_{+,{\rm L}}\\
\psi_{-,{\rm L}}
\end{pmatrix},\;\;
{\cal M}=\begin{pmatrix}
m_{++}&m_{+-}\\
m_{-+}&m_{--}
\end{pmatrix}.\label{Mdef}
\end{equation}
 The condition of particle current conservation is $\sigma_{z}{\cal M}^{\dagger}\sigma_{z}={\cal M}^{-1}$, where the Pauli matrix $\sigma_{z}$ acts on the block structure indicated in Eq.\ \eqref{Mdef}. In the Majorana basis of real mode amplitudes ${\cal M}$ is a real matrix, hence
\begin{equation}
\sigma_{z}{\cal M}^{t}\sigma_{z}={\cal M}^{-1}.\label{MtMinv}
\end{equation}

The transfer matrix has the polar decomposition
\begin{align}
{\cal M}&=\begin{pmatrix}
O_{2}&0\\
0&O_{4}^{\rm t}
\end{pmatrix}\begin{pmatrix}
\cosh\Lambda&-\sinh\Lambda\\
-\sinh\Lambda&\cosh\Lambda
\end{pmatrix}\begin{pmatrix}
O_{3}&0\\
0&O_{1}^{\rm t}
\end{pmatrix}\nonumber\\
&=\begin{pmatrix}
O_{2}&0\\
0&O_{4}^{\rm t}
\end{pmatrix} \exp(-\Lambda\sigma_{x})\begin{pmatrix}
O_{3}&0\\
0&O_{1}^{\rm t}
\end{pmatrix},\label{MOLambda}
\end{align}
where the matrices $O_{p}\in{\rm SO}(2N)$ and $\Lambda={\rm diag}\,(\lambda_{1},\lambda_{2}\ldots,\lambda_{2N})$ are the same as in the polar decomposition \eqref{SOLambda} for the scattering matrix. One readily checks that Eq.\ \eqref{MtMinv} is satisfied.

The polar decomposition \eqref{MOLambda} allows us to compute
\begin{equation}
\sigma_{z}\ln({\cal M}{\cal M}^{\dagger})=\Omega\begin{pmatrix}
0&-2\Lambda\\
2\Lambda&0
\end{pmatrix}\Omega^{\rm t},\;\;\Omega=\begin{pmatrix}
O_{2}&0\\
0&O_{4}^{\rm t}
\end{pmatrix}.\label{lnMM}
\end{equation}
This is an antisymmetric matrix, so it has a Pfaffian,
\begin{equation}
{\rm Pf}\,\bigl(\sigma_{z}\ln{\cal M}{\cal M}^{\dagger}\bigr) = \prod_{n=1}^{2N}2\lambda_{n}.\label{PfMM}
\end{equation}
We have used the identity
\begin{equation}
{\rm Pf}\,BAB^{\rm t}={\rm Det}\,B\,{\rm Pf}\,A,\label{PfBAB}
\end{equation}
with ${\rm Det}\,\Omega=1$.

We conclude that the topological quantum number \eqref{Qdef} can equivalently be written as
\begin{equation}
Q={\rm sign}\left[{\rm Pf}\,\bigl(\sigma_{z}\ln{\cal M}{\cal M}^{\dagger}\bigr)\right].\label{Qprimedef}
\end{equation}

\subsection{How to count Majorana bound states}
\label{howtocount}

To determine the topological quantum number of the disordered superconducting wire we seek the number of Majorana bound states. Particle-hole symmetry ensures that any bound state at zero energy is a Majorana fermion (since the creation and annihilation operators are related by $\gamma^{\dagger}(E)=\gamma(-E)$ and therefore are identical at $E=0$). However, we cannot directly search for zero-energy eigenstates: Even if the Majorana fermions are maximally separated by the entire length $L$ of the wire they will still have a nonzero tunnel coupling which splits their energies apart, away from zero.

The issue here is how to distinguish strongly coupled from weakly coupled Majorana fermions. Any ordinary fermionic excitation, with distinct creation and annihilation operators $a^{\dagger}\neq a$, can be described by a pair of \textit{strongly coupled} Majorana fermion operators $\gamma_{1}=a+a^{\dagger}$, $\gamma_{2}=i(a-a^{\dagger})$. In contrast, the Majorana bound states at opposite ends of the wire are \textit{weakly coupled} Majorana fermions. 

Our geometry of a disordered wire connected at the ends to metal contacts allows for a natural distinction of weak versus strong coupling: We call a pair of Majorana bound states ``strongly coupled'' if they are more strongly coupled to each other than to one of the ends of the wire. Conversely, weakly coupled Majorana bound states are more strongly coupled to one end of the wire than to any other Majorana. The topological quantum number counts only weakly coupled Majorana's.

\begin{figure}[tb]
\centerline{\includegraphics[width=0.9\linewidth]{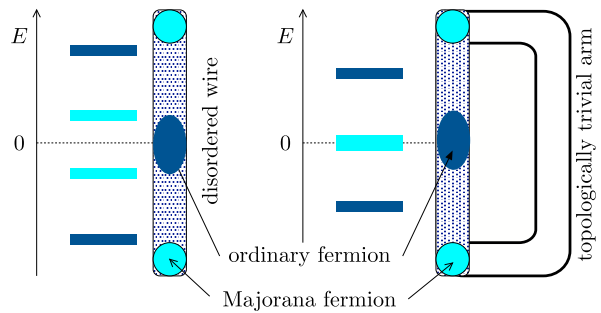}}
\caption{\label{fig_ring}
Procedure to count weakly coupled Majorana bound states in a disordered superconducting wire. Majorana fermions at the two ends of the wire (light blue) are weakly coupled, so their energy is not exactly zero and we need a way to distinguish them from an ordinary fermionic excitation (dark blue). To that end we close the wire into a ring through a topologically trivial superconductor and ask whether destructive interference of the tunnel splitting in the two arms of the ring can produce a pair of two-fold degenerate zero-energy states.
}
\end{figure}

This distinction between weak and strong coupling can be made operational by means of the thought experiment illustrated in Fig.\ \ref{fig_ring}: We close the wire into a ring by connecting the two ends through a superconductor which is in a topologically trivial phase (with a uniform positive gap $\Delta_{0}$). Destructive interference in the two arms of the ring can eliminate the tunnel splitting between a pair of Majorana bound states and produce two-fold degenerate zero-energy eigenstates, if the coupling between the two Majorana's through each arm of the ring is of comparable strength. 

So we vary $\Delta_{0}$ (allowing also for mode mixing at the junction between the two arms of the ring) and find that a number $m$ of two-fold degenerate states appear at zero energy. This means that the disordered wire contains $m$ pairs of Majorana's which are more strongly coupled to the ends of the wire than to each other (otherwise the couplings through the two arms of the ring could not have been equalized by varying $\Delta_{0}$). The number $m$ thus counts the number of weakly coupled Majorana bound states, which gives the topological quantum number $Q=(-1)^{m}$.

\subsection{Topological quantum number of a disordered wire}
\label{derivationPf}

Now that we have an operational definition of the topological quantum number of a finite system, our next step is to relate this to the scattering parameters $\lambda_{n}$ in Eq.\ \eqref{Qdef}. For this purpose it is easiest to work with the transfer matrix, rather than the scattering matrix. An eigenstate $\Psi$ of the ring must be single-valued as we go around the ring, so in terms of the transfer matrices ${\cal M}$ and ${\cal M}_{0}$ of the two arms of the ring we have the condition ${\cal M}_{0}{\cal M}\Psi=\Psi$. This leads to the determinantal condition
\begin{equation}
{\rm Det}\,(1-{\cal M}_{0}{\cal M})=0.\label{DetMM}
\end{equation}

We choose to work in a basis where the orthogonal matrices $O_{p}$ in Eq.\ \eqref{MOLambda} are equal to the unit matrix. Each of the $n=1,2,\ldots 2N$ eigenchannels of the disordered wire can then be treated separately, with $2\times 2$ transfer matrices $M_{n}=\exp(-\sigma_{x}\lambda_{n})$ at zero energy. The topologically trivial arm of the ring (of length $L_{0}$ and coherence length $\xi_{0}=\hbar v_{F}/\Delta_{0}>0$, without any disorder) has transfer matrix $M_{0}=\exp(-\sigma_{x}L_{0}/\xi_{0})$. The condition for an eigenstate at zero energy reads
\begin{equation}
{\rm Det}\,(1-e^{-\sigma_{x}L_{0}/\xi_{0}}e^{-\sigma_{x}\lambda_{n}})=0,\label{Detmm}
\end{equation}
which has a twofold degenerate solution if the ratio $L_{0}/\xi_{0}$ is tuned to the value $-\lambda_{n}$. This is the pair of weakly coupled Majorana bound states in the $n$-th eigenchannel that we are searching for. Because $\xi_{0}>0$, by definition in a topologically trivial phase, the pair exists only if $\lambda_{n}<0$.

We conclude that the number of pairs $m$ of weakly coupled Majorana bound states equals the number of negative $\lambda_{n}$'s, hence
\begin{equation}
Q=(-1)^{m}={\rm sign}\,\left(\prod_{n=1}^{2N}\lambda_{n}\right),
\end{equation}
as announced in Eq.\ \eqref{Qdef}.

\section{Numerical simulations for long-range disorder}
\label{longrange}

Fig.\ \ref{fig_Gthermal} in the main text demonstrates that the quantized thermal conductance at the Majorana phase transition is insensitive  to short-range disorder (correlation length $\xi$ of the order of the lattice constant $a$). Here we show that long-range disorder similarly has no effect on the quantization. (The stability of Majorana bound states against short-range and long-range disorder was investigated in Ref.\ \cite{Lut10b}.)

As before, we solve the scattering problem numerically
by discretizing the Hamiltonian \eqref{HBdG} on a square lattice (with
a total number of $N_\text{tot}$ lattice points in the disordered
region).
The disorder is modeled as a superposition of
impurities with a Gaussian profile,
\begin{equation}\label{gaussian_disorder}
U(\bm{r})=\sum_{i=1}^{N_\text{imp}} U_i \exp\left[-\frac{(\bm{r}-\bm{r}_i)^2}{2 \xi^2}\right],
\end{equation}
where $N_\text{imp}$ is the number of impurities. (We fixed the impurity concentration $n_\text{imp}=N_\text{imp}/N_\text{tot}$ at $5\%$.) The strength $U_i$ of an individual impurity is randomly
distributed in the interval $(-U_0, U_0)$, and
the impurity positions $\bm{r}_i$ are chosen randomly from the
$N_\text{tot}$ lattice points.

\begin{figure}
\includegraphics[width=0.8\linewidth]{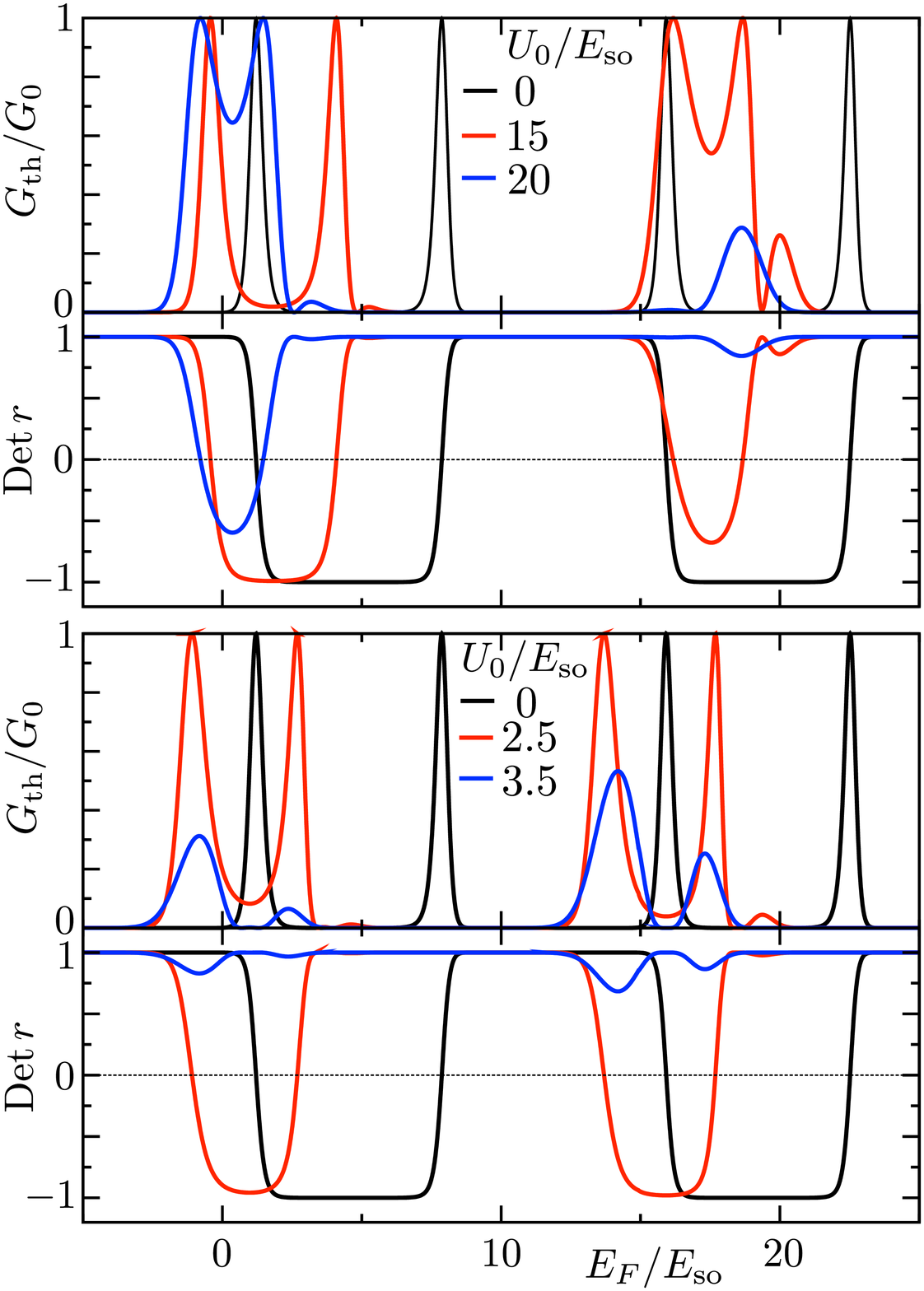}
\caption{Same as Fig.~\ref{fig_Gthermal}, for an impurity potential with correlation length $\xi=2\,a$ (upper panel) and $\xi=10\,a$ (lower panel.}\label{fig_longrange}
\end{figure}

The results of the calculation are shown in Fig.\ \ref{fig_longrange},
for different values of the correlation length $\xi$. In all cases,
we observe as before that the thermal conductance remains quantized as long as
the topological phase persists. For sufficiently strong disorder,
the merging of two peaks signals the disappearance of the topological phase and a breakdown of the conductance quantization.

\section{Electrical conductance and shot noise at the topological phase transition}
\label{GPapp}

The expression \eqref{GPdef} for the nonlocal electrical conductance and shot noise of the superconducting wire can be evaluated further if there is only a single open transmission channel. The $2N\times 2N$ transmission matrix
\begin{equation}
t=\begin{pmatrix}
t_{ee}&t_{eh}\\
t_{he}&t_{hh}
\end{pmatrix}\label{tsubblockdef}
\end{equation}
is then of rank $1$, which means that the $N\times N$ submatrices $t_{ee},t_{hh},t_{he},t_{eh}$ have the dyadic form
\begin{align}
&t_{ee}=|u_{\rm R}\rangle\langle u_{\rm L}|,\;\;t_{hh}=|v_{\rm R}\rangle\langle v_{\rm L}|,\nonumber\\
&t_{he}=|v_{\rm R}\rangle\langle u_{\rm L}|,\;\;t_{eh}=|u_{\rm R}\rangle\langle v_{\rm L}|.\label{tuvdef}
\end{align}
The matrix ${\cal T}_{\pm}$ then becomes
\begin{equation}
{\cal T}_{\pm}=|u_{\rm L}\rangle\langle u_{\rm L}|\bigl(\langle u_{\rm R}|u_{\rm R}\rangle\pm\langle v_{\rm R}|v_{\rm R}\rangle\bigr).\label{Tpmuv}
\end{equation}
Electron-hole symmetry requires $|v_{\rm R}\rangle=|u_{\rm R}^{\ast}\rangle$, hence ${\cal T}_{-}=0$, ${\cal T}_{+}=\frac{1}{2}{\rm Tr}\,tt^{\dagger}$, and thus $G=0$, $P=(e^{3}V_{1}/2h)\,{\rm Tr}\,tt^{\dagger}$.


\begin{thebibliography}{99}
\bibitem{note4} There exist, in addition to class D, four more symmetry classes with a topological phase transition in a wire geometry. As we will show elsewhere, the quantized conductance at the transition point appears generically. This is a manifestation of the ``super-universality'' of Ref.\ \cite{Gru05}.
\bibitem{note3} We do not plot the quantized shot noise peak in a separate figure, because our numerical simulation shows that $P$ in units of $e^{3}V_{1}/2h$ is indistinguishable on the scale of Fig.\ \ref{fig_Gthermal} from $G_{\rm th}$ in units of $G_{0}$.
\end{thebibliography}
\end{document}